\documentstyle[psfig]{mn}

\def\Zycki{$\dot{\rm Z}$ycki}
\def\Rozanska{R\'o$\dot{\rm z}$a\'nska }
\def\Lubinski{Lubi\'nski }
\def\Gierlinski{Gierli\'nski }

\title[Cyg X-1 in the low/hard state]
{On the accretion geometry of Cyg X-1 in the low/hard state}
\author[F. E. Barrio, C. Done \& S. Nayakshin]
    {F. Eugenio Barrio$^1$, Chris Done$^1$ and Sergei Nayakshin$^2$\\
        $^1$ University of Durham, Department of Physics, 
       South Road, Durham DH1 3LE; eugenio.barrio@durham.ac.uk,
chris.done@durham.ac.uk \\
$^2$ Max Planck Institut fur Astrophysik,
Karl-Schwarzschild-Str. 1,
Postfach 1317, D-85741 Garching, Germany
}

\begin{document}

\maketitle

\begin{abstract}

We fit the broad-band RXTE PCA and HEXTE spectrum from 3--200 keV with
reflection models which calculate the vertical ionization structure of
an X-ray illuminated disc. We consider two geometries corresponding to
a truncated disc/inner hot flow and magnetic flares above an
untruncated disc. Both models are able to fit the PCA 3--20 keV data,
but with very different spectral components.  In the magnetic flare
models the 3--20 keV PCA spectrum contains a large amount of highly
ionized reflection while in the truncated disc models the amount of
reflection is rather small. The Compton downscattering rollover in
reflected emission means that the magnetic flare models predict a
break in the spectrum at the high energies covered by the HEXTE
bandpass which is {\em not} seen. By contrast the weakly illuminated
truncated disc models can easily fit the 3--200 keV spectra.

\end{abstract}

\section{Introduction}

The black hole binary (BHB) systems in their low/hard state have
spectra dominated by a power law which rolls over
at $\sim 200$ keV. This continuum is
generally well fit by thermal Comptonization models, in which low
energy photons from the accretion disc are upscattered by energetic
electrons (e.g. the review by Zdziarski 2000). 
To get these hard X-rays, a large fraction of
the gravitational energy released by accretion must be dissipated in
an optically thin environment i.e. not in the disc itself. 
However, there is no consensus on how this happens, or on the geometry
of this hot region. There are currently two main models,
one in which the hot electrons are confined in magnetic
flares above a disc which extends down to the last stable orbit around
the black hole (magnetic flares), and one in which the electrons form
an quasi-spherical hot flow, replacing the inner disc (truncated
disc).

The magnetic flare model is motivated by the discovery that the
physical mechanism for the disc viscosity is a magneto-hydrodynamic
dynamo (e.g. the review by Balbus \& Hawley 2002). Buoyancy could
cause the magnetic field loops to rise up to the surface of the disc,
so they can reconnect in regions of fairly low particle
density, forming a patchy corona. 
Numerical simulations (although these are highly incomplete as in
general the simulated discs are not radiative) do show this
happening (Hawley 2000), but they do not yet carry enough power
to reproduce the observed low/hard state (Miller \& Stone 2000).

The truncated disc model has its physical basis in the accretion flow
equations. The standard Shakura-Sunyaev disc solution assumes that the
accreting material is at one temperature (protons and electrons
thermalize) and that the accretion energy released by viscosity is
radiated efficiently. At low mass accretion rates neither of
these are necessarily true. The thermalization timescale between the
electrons and protons can be long, so the 
flow is intrinsically a two temperature plasma. Where the electrons
radiate most of the gravitational energy through Comptonization of
photons from the outer disc then the hot inner flow is given
by Shapiro, Lightman \& Eardley (1976). Alternatively, if the
protons carry most of the accretion energy into the black hole then
this forms the advection dominated accretion flows (Narayan \& Yi
1995). These are related, as in general both advection
and radiative cooling are important for a hot accretion flow
(Chen et al. 1995; Zdziarski 1998).

These two models of the accretion flow have very different geometries.
A potential way to test the geometry is with X-ray reflection.  The
amount of reflection scales with the solid angle subtended by the
optically thick material while the relativistic smearing of the atomic
features shows how far this material extends into the gravitational
potential (Fabian et al.  2000). With magnetic flares, the disc
subtends a solid angle of $2\pi$ as seen from the X-ray source, and
extends down to the last stable orbit. Its reflected spectrum should
be large and strongly smeared.
Conversely, a truncated disc illuminated by an inner hot flow
subtends a solid angle $\le 2\pi$, and the reflected spectrum is small
and is only weakly smeared by relativistic effects.

The BHB spectra in the low/hard state show overwhelmingly that the
solid angle is significantly less than $2\pi$, and that the smearing
is less than expected for a disc extending down to the last stable
orbit (\Zycki\ Done \& Smith 1997; 1998; 1999; \Gierlinski\ et
al. 1997; Done \& \Zycki\ 1999; Zdziarski et al. 1999; Gilfanov,
Churazov \& Revnivtsev 1999; 2000).  While this is clearly consistent
with the idea that the disc is truncated in the low/hard state, the
magnetic flare models can be retrieved in several ways.

Firstly, magnetic reconnection on the Sun is known
to produce an outflow in the coronal mass ejection events. In the
extreme conditions close to the black hole it is possible that this
outflow velocity could be large so that the hard X--ray radiation is
beamed away from the inner disc (Beloborodov 1999).  

An alternative explanation for the lack of reflection and smearing is
that the inner disc or top layer of the inner disc is completely
ionized. There are then no atomic features, and the disc reflection is
unobservable in the 2--20 keV range as it appears instead to be part
of the power law continuum (Ross \& Fabian 1993; Ross, Fabian \& Young
1999).  However, these models with passive illumination of the disc
require a fairly sharp transition between the extreme ionization and
mainly neutral material (Done \& \Zycki\ 1999; Done, Madejski \& \Zycki\
2000; Young et al. 2001). Such a transition can be produced as the
disc {\em responds} to the intense X--ray illumination. There is a
thermal ionization instability which affects X--ray illuminated
material in pressure balance, which can lead to a hot, extremely
ionized skin forming on top of the rest of the cooler, denser, mainly
neutral disc material (Field 1965; Krolik, McKee \& Tarter 1981;
Kallman \& White 1989; Ko \& Kallman 1994; \Rozanska\ \& Czerny 1996;
Nayakshin, Kazanas \& Kallman 2000, hereafter NKK; \Rozanska\ \&
Czerny 2000; Nayakshin \& Kallman 2001; Ballantyne, Ross \& Fabian
2001). Such X-ray illuminated disc models can fit the 2--20 keV data
from BHB with reflection from a disc which subtends a solid angle of
$2\pi$ and extends down to the last stable orbit around a black hole
(Done \& Nayakshin 2001b).

While the truncated disc and magnetic flare/X-ray illuminated disc
models are indistinguishable with current data in the 2--20 keV range,
they are very different at higher energies. In the magnetic flare
models, a lot of the 2--10 keV 'continuum' is actually ionized
reflection, so the true continuum level is lower than in 
the truncated disc models, in which reflection is small. At higher
energies, where reflection is negligible, the lower continuum
level for the magnetic flare models leads to a smaller predicted flux
in the 100-200 keV range than for the truncated discs (Done \& Nayakshin
2001a). Here we fit both truncated disc and magnetic flare/X-ray
illuminated disc models to the 3--200 keV PCA/HEXTE spectrum of the
hard/low state spectra of Cyg~X-1. We show that the
truncated disc models provide an excellent fit to these data, but that
the magnetic flare/X-ray illuminated disc does not, as it dramatically 
underpredicts the 100--200 keV spectrum. 

Similar conclusions were independently reached by Maccarone \& Coppi
(2002) from fits to the Cyg X-1 broad-band spectrum. However, they
approximated the reflection from magnetic flares by highly ionized,
single zone reflection models rather than the full complex ionization
reflection models used here. Here we are able to show explicitally
that the complex ionization magnetic flare models do not fit the high
energy spectrum, while the truncated disc models do. 

\section{Models}

We use the complex ionization reflection code, {\sc xion}, described in NKK
which computes the self-consistent vertical ionization
structure of an X-ray illuminated disc at a given radius.  The
reflected and diffuse (line and recombination continua) emission from
this are smeared by the relativistic effects
expected from a disc, and then these spectra are summed over all
radii to get the total disc reflected emission for any given
source/disc geometry. These models have been tabulated for use in
{\sc xspec} for a central sphere/truncated disc geometry, assuming that the
hot inner flow has emissivity $\propto r^{-3}$, and starts at the
radius where the optically thick disc truncates (Nayakshin et al.
2002, in preparation). 
They have also been tabulated for a magnetic flare geometry,
which results in much stronger illumination of the disc and so gives a
deeper ionized skin (Nayakshin 2000). With magnetic flares, the
luminosity of the flares is assumed to scale as the local disc flux,
so that $f_x/f_{disc}$ remains constant with radius. These tabulated
models are hereafter referred to as {\sc xion-disc} (truncated disc)
and {\sc xion-flares} (magnetic flares). The free parameters are the
mass accretion rate (this determines the unilluminated disc structure
i.e. the starting density as a function of height), $f_x/f_{disc}$
which determines the depth of the ionized skin, and the shape of
the illuminating spectrum ($\Gamma$ and $E_{cut}$ assuming an
exponentially cutoff power law). For the truncated disc models there
is also the inner radius of the accretion disc, $R_{in}$, which
controls the amount of relativistic smearing of the iron line
features.

The {\sc xion} code includes Compton up and downscattering by electrons in
the disc. Compton upscattering can be very important in smearing the
iron features in ionized discs (Ross et al. 1999), while
Compton downscattering determines the high energy rollover in the
reflected spectrum (Lightman \& White 1988).  However, the
approximations used for Compton scattering in the {\sc xion} code become
progressively less accurate at higher energies (NKK)
The Appendix discusses how we merge the more accurate high
energy Compton downscattering of the standard {\sc xspec} {\sc pexriv} based
reflection models (Magdziarz \& Zdziarski 1995) with the {\sc xion} results.

\section{Spectral fitting}

We used the hardest spectra seen from multiple observations of the Cyg
X-1 hard/low state (Gilfanov et al. 1999).
We extracted the data using the {\sc Rex} data analysis
script with the bright source background from the top layer of PCA
detectors 0 and 1. Previous work has shown that this configuration
gives a good fit to the Crab data with 0.5 per cent systematic error
for RXTE Epoch 3 data (Wilson \& Done 2001).  We extracted the
simultaneous HEXTE data from cluster 0. In fits with both the PCA and
HEXTE data we allow a normalisation offset between the two
instruments to take into account the cross-calibration
uncertainties. We use {\sc xspec} version 11 (Arnaud et al 1996) and quote
all error bars as $\Delta\chi^2=2.7$, corresponding to 90 per cent
confidence limits on 1 parameter, and fix the absorbing column to
$N_H=6\times 10^{21}$ cm$^{-2}$ (Ba{\l}uci{\'n}ska-Church et al. 1995).

First we fit the PCA data alone with the old, single ionization
parameter reflection models so that we can compare the new reflection
models with previous fits.  We use the {\sc rel-repr} model of \Zycki\
et al. (1998), which is based on the publically available {\sc
pexriv} reflection but includes the self consistently produced iron
line emission and relativistic smearing. We assume a power law
continuum model, and also include neutral, unsmeared reflection which
can arise from the companion star or outer accretion disc (Ebisawa et
al 1996).  This gives a $\chi^2_\nu=30/39$ and the amount of
reflection from the accretion disc is low, with $\Omega/2\pi=0.13\pm
0.08$, implying a truncated disc.

We then replace the single ionization accretion disc reflection with
the {\sc xion-disc} model.  The simplest version of this truncated
disc geometry assumes the hot inner source is spherical, and starts
where the cool disc truncates. This results in a model in which
$\Omega/2\pi$ is fixed at $\sim 0.3$, rather larger than inferred from
the simple reflection models above, and so gives a somewhat larger
$\chi^2_\nu=53/40$. If instead we allow the amount of reflection to be
smaller than this (e.g. if the hot source has a flattened rather than
spherical geometry) then we get $\chi^2_\nu=43/39$ for $f_x/f_d=6$.
This is a worse fit by $\chi^2$ of $13$ than that for the single
ionization zone reflection models ({\sc rel-repr}, above). 
The two models give different fits as even a weakly
illuminated disc produces a low optical depth ionized skin. This
reflects some small fraction of the incident continuum at low energies
(see Fig. A1 in the appendix), changing the curvature of the
reflected continuum.

We then use {\sc xion-flare} to model the magnetic flare geometry in
which $\Omega/2\pi$ is fixed at 1 and the inner disc extends down to 3
Schwarzchild radii. This gives $\chi^2_\nu=47/41$ (both the solid
angle of reflection and the inner disc radius are fixed, leading to 2
fewer degrees of freedom). This confirms the results of Done \&
Nayakshin (2001b) that magnetic flares can give as good a fit to the
data below 20 keV as a truncated disc. The PCA data are consistent
with the presence of a large amount of ionized reflection, i.e. with a
disc which subtends a large solid angle to the X-ray source and
extends down to the last stable orbit.

However, these two very different reflection models should give very
different spectra at higher energies. Reflection {\em must} rollover
at high energies due to Compton downscattering in the disc.  Fig. 1
shows the truncated disc (upper panel) and magnetic flare (lower
panel) PCA fits extrapolated to the HEXTE data.  The truncated disc
model has little reflection, so there is little rollover in the total
spectrum. By contrast, the magnetic flare models have a lot of highly
ionized reflection, so the contribution of the Compton downscattering
rollover is important. Plainly there is no such rollover in the HEXTE
data, and the magnetic flare description of the PCA data strongly
underestimates the higher energy spectrum. This is made even more
marked by the fact that a power law is an overestimate of the high
energy continuum: the true Comptonized continuum has a thermal cutoff
at 150-200 keV (\Gierlinski\ et al. 1997). By contrast, the truncated
disc models can qualitatively describe the high energy data.

\begin{figure}
\psfig{file=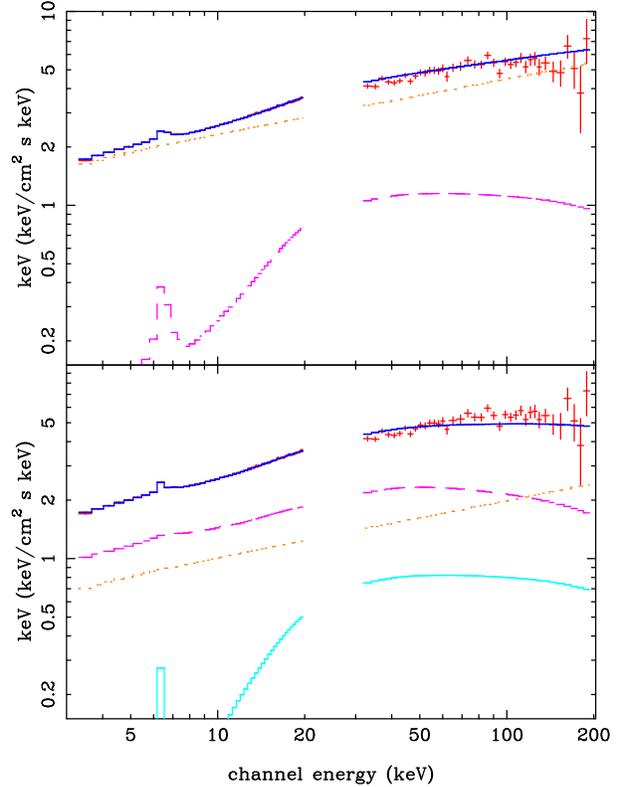,width=0.45\textwidth}
\caption{ 
The upper panel shows the PCA fit to a power law and
truncated disc
model extrapolated through the HEXTE data, while the
lower panel shows this for the magnetic flares. Both figures have the
power law continuum as a dotted line, while the dashed line is the
reflected component from the accretion disc. The intrinsic continuum
level for the truncated disc is much higher than for magnetic flares
as the magnetic flares have a large contribution of reflected emission
in the PCA. The reflected continuum has a {\em rollover} at $\sim 100$
keV, so the extrapolated high energy spectrum is dominated by the
intrinsic continuum and the magnetic flare model predicts a much
smaller HEXTE flux than the truncated disc. Both fits have comparable
$\chi^2$ in the PCA, but the magnetic flares underpredict the high
energy data while the truncated disc model matches fairly well. Both
fits include neutral, unsmeared reflection from the companion
star/outer disc (lower solid line in the bottom panel) but is
too small to be seen in the truncated disc model.  
}
\end{figure}

While this argues strongly against a large ionized reflection
component in the data, true spectral fitting is required to show
conclusively that this is the case. We do joint fits of the PCA and
HEXTE data, using the Comptonization model, {\sc compps}, of Poutanen
\& Svensson (1996) rather than a power law for the continuum.  With
reflection from a spherical source within a truncated disc (and
neutral, unsmeared reflection from the companion star) then this gives
$\chi^2_\nu=114/77$, while allowing the reflection amount to vary (as
in a flattened source) gives $\chi^2_\nu=86/76$. We show the unfolded
spectrum and residuals from this model in Fig. 2, where the continuum
parameters are $\tau\sim 0.75$ and $kT_e=200$ keV.  By contrast, using
reflection from magnetic flares gives a very poor fit, with
$\chi^2_\nu=259/78$ (Fig. 3) for a continuum with $\tau\sim 0.75$ and
$kT_e\sim 210$ keV.

\begin{figure}
\psfig{file=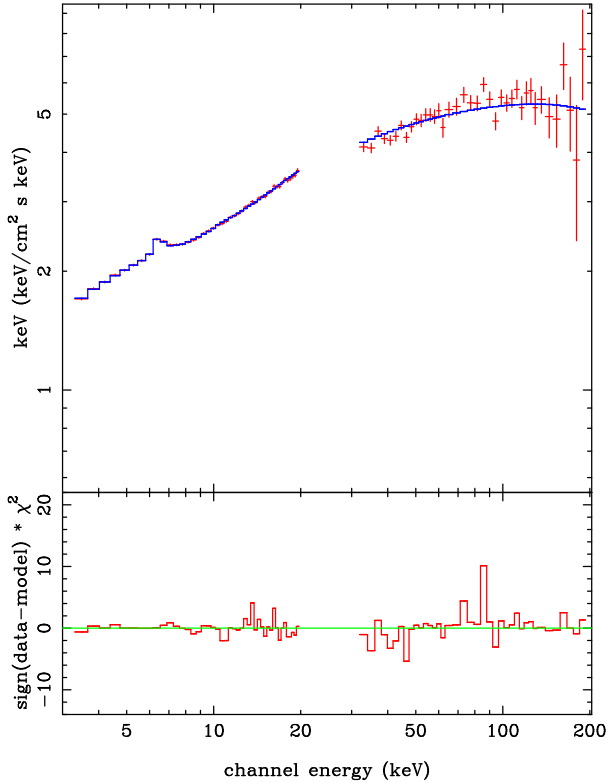,width=0.45\textwidth}
\caption{The PCA and HEXTE data fit to a Comptonized continuum plus
illuminated truncated accretion disc model {\sc xion-disc}, plus
neutral, unsmeared reflection from the companion star/outer disc.
The upper panel shows the unfolded spectrum while the lower panel
shows residuals to the fit.
}
\end{figure}

\begin{figure}
\psfig{file=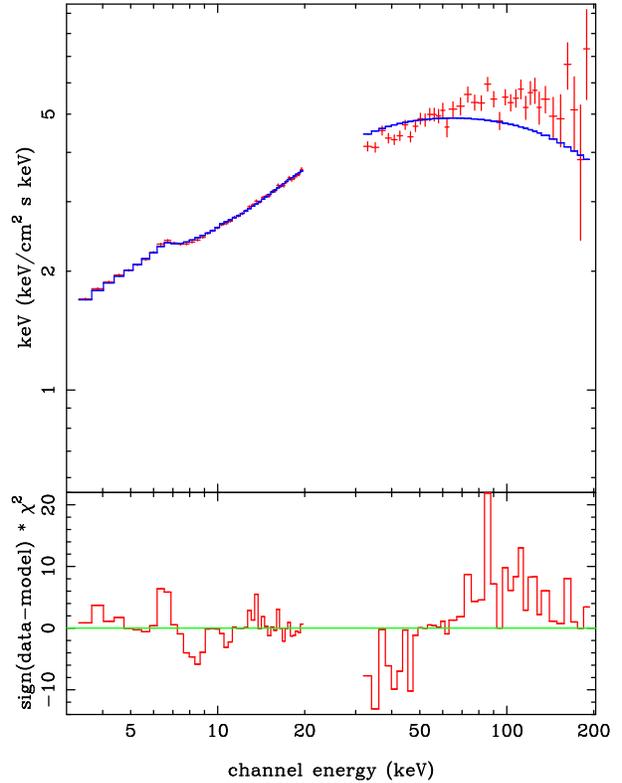,width=0.45\textwidth}
\caption{As for Fig. 2, but with the {\sc xion-flare} model for
reflection from an X-ray illuminated untruncated disc. The 
large amount of reflection with its rollover at $\sim 100$ 
keV means that the slope of the HEXTE spectrum 
is predicted to be considerably steeper than that of the PCA, while the 
observed HEXTE data have a rather similar slope.
}
\end{figure}

All these fits assumed isotropic seed photons for the Compton
scattering. However, this is unlikely to be the case since the seed
photons come from the disc. While there is some range of incident
angle for the truncated disc geometry, the magnetic flares are
illuminated preferentially from below. For a Comptonizing plasma with
optical depth of order unity and temperature of $\sim 200$ keV
(Sunyaev \& Titarchuk 1980; \Gierlinski\ et al. 1997) then this seed
photon anisotropy can have a marked effect on the spectrum (Haardt \&
Maraschi 1993; Poutanen \& Svensson 1996).  We use the {\sc compps}
code in a slab geometry, where the hot plasma is illuminated from
below (Poutanen \& Svensson 1996) as an approximation to the magnetic
flare geometry.  We refit the data with the magnetic flare reflection
model using this new continuum.  This gives an even worse fit than
before, with $\chi^2=531/78$, where the continuum has $\tau\sim 1$ and
$kT_e=85$ keV.  The reason the fit is so poor is that the seed photon
anisotropy leads to a break in the spectrum, such that the low energy
spectrum is harder than the high energy spectrum (Haardt \& Maraschi
1993; Haardt et al. 1993). The low energy spectrum then predicts even
less high energy emission than for isotropic seed photons.
Fig. 4 shows the unfolded spectrum and residuals from this model.

\begin{figure}
\psfig{file=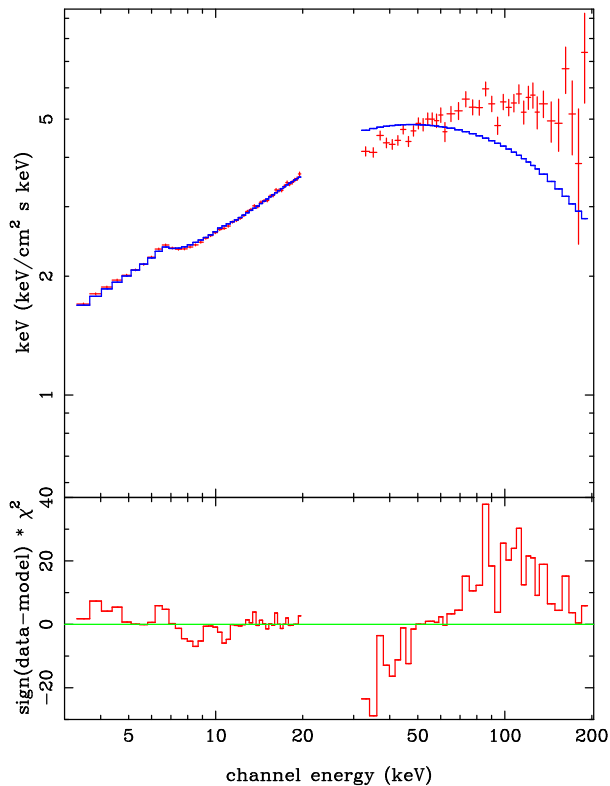,width=0.45\textwidth}
\caption{
As for Fig. 3 (reflection from {\sc xion-flare}), 
but with the anisotropic Compton continuum 
expected from X-ray emission regions above an accretion disc.
The fit is even worse than for isotropic emission as the Compton
continuum has a break to a steeper spectrum at high energies due to
the anisotropic illumination of the seed photons from the accretion
disc. Note the change in scale for the $\chi^2$ panel compared to
Figs. 2 and 3.
}
\end{figure}

\section{Discussion}

The data clearly show that the 2--20 keV spectrum from the low/hard
state of Cyg X-1 does {\em not} contain a large fraction of highly
ionized reflection. This rules out models
which have static magnetic flares above an untruncated disc unless the
flares have a spectrum which is much harder than that predicted by a
single temperature Comptonization model. 

One way to get some spectral hardening which is expected in a magnetic
flare geometry but is neglected in our modelling is for the flares to
comptonize some fraction of the reflected photons.  The reflected
photons are from the disc, so are intercepted by the hot electrons in
the flares in the same way as the soft seed photons, and are Compton
upscattered to form a hard continuum.  With flares covering most of
the disc then this can make a $50-100$ per cent increase to the flux
at 200 keV (Petrucci et al. 2001). A covering fraction of unity for
the flares (slab corona geometry) is normally ruled out for the
low/hard spectra as this produces many non-reflected, thermalized
photons from the hard X-ray illumination. These are also Comptonized
by the corona, leading to spectral indices which are too soft to
explain the low/hard state (Pietrini \& Krolik 1995; Stern et
al. 1995; Zdziarski et al. 1998).  However, a covering fraction of
$\sim 0.5$ might give enough Compton scattering of reflection to
produce the required $\sim 50$ per cent excess emission at 200 keV,
while also allowing enough seed photons to escape to produce the
required hard continuum.

The problems that the magnetic flare models have in matching the high
energy flux are exacerbated by the anisotropy break which should be
present in the continuum in this assumed geometry, but which has never
been convincingly observed (\Gierlinski\ et al. 1999).  If the
magnetic flares are to fit the high energy data then the anisotropy
break must be hidden by having a multiple temperature Comptonised
continuum. At some level there {\em must} be a distribution of
electron temperatures: it is almost inconceivable that a single
temperture distribution can be maintained, especially as the 
sources are variable so the flare spectra
probably evolve with time (e.g. Poutanen \& Fabian 1999). A multiple
temperature continuum gives another way to boost the high energy flux
so that we do not necessarily need to strongly Comptonise the
reflected continuum. However, it still seems somewhat contrived that a
combined complex continuum plus complex ionization reflection spectrum
should so precisely mimic a simple single temperature continuum,
truncated disc reflection.

By contrast, a truncated disc/hot inner flow geometry at low mass
accretion rates is compatible with the observed hard continuum, lack
of anisotropy break, low amount of reflection and relativistic
smearing.  It can also explain the low temperature and luminosity of
the direct emission from the disc (e.g. Esin et al. 2000). This
geometry can also give a qualitative explanation for a range of
observed correlations if the truncation radius decreases with
increasing (average) mass accretion rate.  The disc penetrates further
into the hot flow, increasing the seed photon flux intercepted by the
hot inner flow, leading to a softer continuum spectra.  This changing
geometry gives a larger solid angle subtended by the disc, leading to
an increasing amount of reflection (Poutanen Krolik \& Ryde 1997;
Zdziarski et al. 1999; Gilfanov et al. 1999; 2000), and relativistic
smearing (\Zycki\ et al. 1999; Gilfanov et al. 2000; \Lubinski\ \&
Zdziarski 2001). The variability power spectra are also affected as
they contain characteristic frequencies which are most probably linked
to the inner edge of the disc, so this can explain the correlated
increase in break and quasi-periodic oscillation frequency (e.g. the
review by van der Klis 2000; Churazov, Gilfanov \& Revnivtsev 2001).
Lastly, the collapse of an inner hot flow when it becomes optically
thick gives a physical mechanism for the state transition (Esin
McClintock \& Narayan 1997). 

Thus, if one were to choose between relatively straightfoward models, then
the truncated disk is definitely favored by our analysis while the
magnetic flare model is ruled out. However, the straightforward solutions
may be too simple to describe the complexity of accretion disk structure
near the black hole. The role and magnitude of secondary effects
(comptonization of the reflection component; multi-temperature nature of
flares) not taken into account in our modelling needs to be clarified in
the future with detailed calculations. 

{}

\appendix

\section{A comparison of {\sc xion} and {\sc pexriv} reflection}

The main objective of the {\sc xion} code is to accurately calculate
the vertical ionisation structure of the X-ray illuminated disc, so
the atomic physics/photoionisation calculations are treated in great
depth while compton scattering is approximated in the same manner as
in Ross \& Fabian (1993).  These approximations are adequate at lower
energies, where the energy shifts from compton scattering are small,
but become progressively worse at higher energies where downscattering
is large. By contrast, the standard {\sc
pexriv} reflection model in XSPEC treats compton scattering very
carefully (Magdziarz \& Zdziarski 1995), but has very crude ionisation
balance and does not include any vertical structure (Done et
al. 1992).

The two codes should give comparable results when the vertical
structure is unimportant i.e. when the X-ray illumination is very
small. Fig. A1 shows a comparison of the two codes with {\sc
xion-flares} at $f_x/f_d =0.02$ and $\dot{m}=0.001$ (the lowest
tabulated values) for a magnetic flare geometry at inclination of
$30^\circ$. The dotted line shows the comparable {\sc pexriv} results
($\Omega/2\pi=1$, $\xi=0$, inclination of $30^\circ$). The continuum
for both is an illuminating power law of index $\Gamma=1.7$ with
exponential rollover at 300 keV. The differences are obvious.  At low
energies these are expected as the {\sc xion} code has a residual
ionized skin in even a weakly illuminated disc. One would have to
extend the grid in {\sc xion} to even lower values of the ``gravity
parameter'' to render effects of the skin completely negligible in the
soft X-ray range.  The ionisation effects should nevertheless be very
small at high energies where the two codes should yield very simillar
results.  However the Compton hump extends to much higher energies in
{\sc pexriv} than in {\sc xion}, and the normalisation is different by
a factor of 1.3 at 20 keV. 

The difference in normalization is due mainly to the difference in
illumination law used.  Magdziarz \& Zdziarski (1995) have specific
intensity $\propto 1/{\rm cos} \theta$, while {\sc xion} assumes a
single ray, incident onto the surface at $45^\circ$. The first form of
the illumination law is appropriate for an optically thin {\em full}
corona (with a covering fraction equal to unity) which is now known
not to work for Cyg X-1 (Gierlinski et al. 1997); the use of a fixed
value of $\theta=45^\circ$ in {\sc xion}, on the other hand, stems
from the necessity to keep the {\sc xion} runtime to a manageable
minimum. Thus both illumination laws are not expected to be strictly
correct in reality and are approximations good to some 10-20 \%. Also,
{\sc xion} is a table model so linear interpolation from the nearest
tabulated parameter values can introduce some inaccuracy.

The deficit of photons at $\sim 200$ keV is a more serious problem.
It is mainly due to the energy grid of {\sc xion} only extending to
200 keV, again from the need to keep the {\sc xion} runtime to a
managable level. However, it is these high energy continuum photons
(which are included in {\sc pexriv})
which are downscattered to form the reflected continuum above 30 keV.
Thus {\sc xion} as tabulated here gives
too few reflected photons at high energies compared with {\sc pexriv}.

\begin{figure}
\psfig{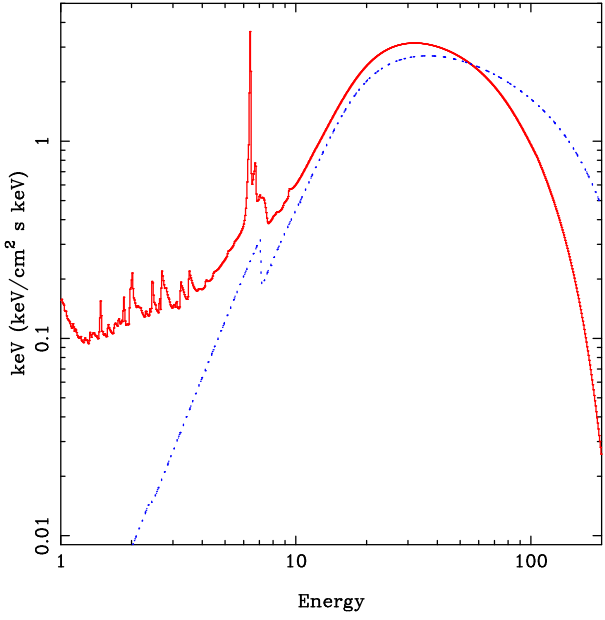}
\caption{ The solid line shows the lowest ionisation {\sc xion}
reflected spectrum from a power law continuum with $\Gamma=1.7$, with
exponential rollover at 300 keV. We use the {\sc xion-flare} model, so
the disc subtends a solid angle of $2\pi$, and assume an inclination
of $30^\circ$. The dashed-dotted line shows the reflected spectrum
calculated under the same conditions using the {\sc pexriv} model.
{\sc xion-flare} predicts much more reflected flux at low energies as
even at the lowest tabulated ionization there is still a residual, low
optical depth, ionized skin which reflects a small fraction of the
continuum.  The difference in shape at high energies and the slight
difference in normalisation at 20-30 keV is due to the approximate
treatment of compton downscattering in the {\sc xion} code and the
different assumed illumination.  }
\end{figure}

An ideal model would accurately calculate {\em both} the ionization
structure and compton scattering for a broad range of parameters and
do it {\em fast}. Since such model is not feasible due to computer
limitations, we use the ionization structure of {\sc xion} at low
energies ($< 20$ keV) and the compton downscattering calculations of
{\sc pexriv} at high energies ($> 30$ keV), having matched the
normalisation of the {\sc pexriv} reflection to that of {\sc xion} at
20 keV. The gap in good data between 20--30 keV means that the two
different models can be used for the PCA and HEXTE data, respectively,
rather than having to interpolate between them.

For a single temperature compton continuum fit to the data above 10
keV in this approach with the {\sc xion-disc} model gives
$\chi^2_\nu=62/59$.  By comparison, the reflection model built into
the {\sc compps} code (which is based on {\sc pexriv}) has
$\chi^2_\nu=65/59$ showing that the matching condition is adequate.

\end{document}